\documentclass{article}
\usepackage{color}
\usepackage{indentfirst}
\usepackage{a4wide}
\usepackage{epsfig}
\usepackage{graphics}
\newcommand{\R}{\mbox{$I \kern -4pt R$}}             

\newtheorem{theorem}{Theorem}

\newtheorem{corollary}[theorem]{Corollary}

\newtheorem{lemma}[theorem]{Lemma}

\newenvironment{proof}[1][Proof]{\noindent\textbf{#1.} }{\ \rule{0.5em}{0.5em}}
\begin{document}

\title{Bifurcations in a convection problem with temperature-dependent
viscosity}
\author{F. Pla$^1$, H. Herrero$^1$ and O. Lafitte$^2$ \\
{$^1$ {\small Departamento de Matem\'aticas}} {\small Facultad de
Ciencias Qu\'{\i}micas}\\
{\small Universidad de Castilla-La Mancha} {\small 13071 Ciudad Real, Spain.}\\
{\small Phone: 34 926 295412, e-mail: Francisco.Pla@uclm.es, Henar.Herrero@uclm.es}\\
{$^2$} {\small Institut Galil\'ee (Universit\'e Paris 13)} \\
{\small 99 av. Jean-Baptiste Cl\'ement 93430 Villetaneuse,} \\
{\small and DM2S, CEA Saclay 91191 Gif sur Yvette Cedex, France.} \\
{\small Phone: 0149403570, e-mail: lafitte@math.polytechnique.fr}}
\maketitle
\begin{abstract}
A convection problem with  temperature-dependent viscosity in
 an infinite layer is presented. As described, this problem
has important applications in mantle convection. The existence of
a stationary bifurcation is proved together with a condition to
obtain the critical parameters at which the bifurcation takes
place. For a general dependence of viscosity with temperature a
numerical strategy for the calculation of the critical bifurcation
curves and the most unstable modes has been developed. For a
exponential dependence of viscosity on temperature the numerical
calculations have been done. Comparisons with the classical
Rayleigh-B\'enard problem with constant viscosity indicate that
the critical threshold decreases as the exponential rate parameter
increases.
\end{abstract}

\section{Introduction}
A classical subject in fluid mechanics is the problem of
thermoconvective instabilities in fluid layers driven by a
temperature gradient \cite{benard,pearson,batchelor}. Gravity
and/or capillary forces are responsible for the onset of motion
when the temperature difference exceeds a certain threshold. The
existence of a stationary bifurcation  for the gravity driven case
(the classical Rayleigh-B\'enard problem) has been rigorously
proved in \cite{proof1,ma}. Above the threshold convective
solutions consist of rolls \cite{lega}. Most studies on the
Rayleigh-B\'enard problem consider a constant viscosity
\cite{mer2,tcfd,park}, although interest on convection problems
with temperature dependent viscosity has increased \cite{torrance,
booker, richter, laure,ke,moresi,trompert} due to the fact that
this dependence is a fundamental feature of mantle convection. The
formation of rocks indicates that viscosity in the interior of the
Earth strongly depends on temperature, and this influences the
thermal evolution of the mantle \cite{kincaid,davies}. Some
theoretical or numerical studies with temperature dependent
viscosity in finite boxes can be found for instance in Refs.
\cite{torrance, ke, moresi,manga}. In \cite{severin} several types
of dependence have been used, but the most common is an
exponential dependence \cite{moresi,ke}.\par This paper plans a
theoretical and numerical study of the instabilities in a gravity
driven convection problem with viscosity dependent on temperature.
The domain, which is finite in the vertical coordinate and
infinite in the horizontal plane, contains a viscous fluid which
is heated uniformly from below.
 In this work, for a general dependence of viscosity on temperature,
 we prove the existence of a stationary
bifurcation as the temperature gradient increases. And a condition
to obtain the critical parameters at which the bifurcation takes
place has been found. For a viscosity with an exponential
dependence on temperature both the critical bifurcation curves and
the most unstable modes have been numerically computed.
Comparisons with the classical Rayleigh-B\'enard problem with
constant viscosity indicate that the critical threshold decreases
as the exponential rate parameter increases. We find the critical
bifurcation curves for this case as a function of the exponential
parameter.\par The paper is organized as follows. In section 2 the
formulation of the problem with the equations and boundary
conditions (bc) and the conductive solution are presented. Section
3 develops the linear stability of the conductive solution and
deduces the theoretical results on the existence and conditions of
bifurcation. In section 4 numerical results  on eigenmodes and
critical curves are presented. Finally in Section 5 conclusions
are detailed.

\section{Formulation of the problem}

\bigskip The considered physical set-up is shown in figure 1. A fluid layer
of depth $d$ ($z$ coordinate) is placed between two parallel
boundless plates. On the bottom
plate a temperature $T_o$ is imposed and on the upper plate a temperature $%
T_{1}=T_{o}-\Delta T=T_{o}-\beta d$, where $\beta$ is the vertical
temperature gradient. The governing equations are the continuity
equation
\begin{equation}
\nabla \cdot \vec{v}=0,
\end{equation}%
the Navier-Stokes equations
\begin{equation}
\partial _{t}\vec{v}+\left( \vec{v}\cdot \nabla
\right) \vec{v}=-\frac{\rho }{\rho _{o}}g\vec{e_{3}}-%
\frac{1}{\rho _{o}}\nabla P+div(\nu (T)\cdot (\nabla \vec{v}+
(\nabla  \vec{v})^t)),
\end{equation}%
and the heat equation,%
\begin{equation}
\partial _{t}T+\vec{v}\cdot \nabla T=K\Delta T.
\end{equation}%
Here $\vec{v}=\left( v_{x},v_{y},v_{z}\right) \left(
x,y,z,t\right) $ is the velocity vector field, $P$ is the
pressure,
 $T$ the temperature field, $\rho $ is the density, which
dependence on temperature is assumed to be $\rho =
\rho_{1}(1-\alpha \left( T-T_{1}\right))$, being $\rho _{1}$ the
mean density at temperature $T_{1}$, $\alpha $ is the thermal
expansion, $K$ is the thermal diffusion, $g$ is the acceleration
of the gravity and $\vec{e}_3$ is the unitary vector in the
vertical direction. $\nu (T)$ is the viscosity that is assumed to
be a positive, convex and bounded function of temperature $T$.
\par
Now, the boundary conditions are introduced. The temperature on
the bottom plate, $z=-d/2$, and on the upper plate, $z=d/2$, are
respectively
\begin{equation}
T\left( z=-\frac{d}{2}\right) =T_{o},\,\ \ T\left(
z=\frac{d}{2}\right) =T_{1}.
\end{equation}
The boundary conditions for the velocity correspond to rigid,
no-slip at the bottom plate and non-deformable and free-slip at
the upper plate. They are expressed as follows,
\begin{equation}
v_x = v_y = v_z =0\,\ \mbox{on }z= -\frac{d}{2},\;\;\partial_z v_x
= \partial_x v_y = v_z = 0\,\ \mbox{on }z= \frac{d}{2}.
\end{equation}
\bigskip
The simpler solution of the hydrodynamic equations is the
conductive solution. For that solution the temperature only
depends on the vertical component and the fluid is at rest,
\begin{equation}
\partial_x T=\partial_y T=\partial_t T=0,\;\vec{v}=\vec{0}.
\end{equation}%
Under these assumptions the hydrodynamic equations
become,
\begin{eqnarray}
0&=&-\rho g\vec{e_{3}}- \nabla P,  \\
 0&=&K\Delta T,
\end{eqnarray}
which solution is the conductive state
\begin{eqnarray}
 T_{b}(z)&=&T_{o}-\beta \left( z+\frac{d}{2}\right), \\
P_b\left( z\right) &=&P_{o}-\rho _{o}g\left[ z\left(
1+\frac{\alpha \beta d}{2}
\right) +\frac{\alpha \beta z^{2}}{2}\right], \\
\vec{v}_b&=&\vec{0},
\end{eqnarray}
where $P_0$ is an arbitrary constant.
\section{Linear stability of the conductive solution}
 Now, we are interested
in the study of the linear stability analysis for the conductive
solution. Therefore, we perturb it as follows,%
\begin{eqnarray}
 T(x,y,z,t)&=&T_{b}(z) +\theta (x,y,z,t),\\
P\left( x,y,z,t\right) &=&P_{b}(z)+\delta P(x,y,z,t), \\
\vec{v}(x,y,z,t)&=&\vec{v}_b +\vec{u}(x,y,z,t),
\end{eqnarray}
and introducing these expressions into the hydrodynamic equations
the following equations for the perturbation fields are obtained,
\begin{eqnarray}
&& \nabla \cdot \vec{u}=0, \nonumber\\
&& \partial _{t}\vec{u}+\left (\vec{u}\cdot\nabla\right )\vec{u}=\alpha \theta g \vec{e}_3-\frac{1}{%
\rho _{o}}\nabla \left( \delta P\right) +\mbox{div}(\nu (T)\cdot (\nabla \vec{u} +
(\nabla \vec{u})^t)),\label{perturbaciones} \\
&& \partial _{t}\theta+\left (\vec{u}\cdot\nabla\right )\theta
=K\Delta \theta +\beta u_{z}. \nonumber
\end{eqnarray}%
Taking into account the following expressions,
\begin{eqnarray}
&& \nu(T) = \nu(T_b+\theta)=\nu(T_b) + \nu'(T_b) \theta +
O(\theta^2), \nonumber \\
&& \nu(T_b) = \nu(T_0 - \beta (z+d/2))=\nu(z), \nonumber \\
&& \mbox{div}(\nu (T)\cdot (\nabla \vec{u} + (\nabla \vec{u})^t))=
 \tilde{L} \vec{u} + N(\vec{u},\theta), \nonumber
\end{eqnarray}
where $\tilde{L}\vec{u}$ is the linear part and
$N(\vec{u},\theta)$ includes the nonlinear terms as follows
\begin{eqnarray}
\tilde{L}\vec{u}&=&\mbox{div}\left(\nu (z)\cdot \left
(\nabla\vec{u}+(\nabla \vec{u})^t\right)\right),\nonumber\\
N(\vec{u},\theta)&=&\mbox{div}\left( \left(\nu'(T_b) \theta +
O(\theta^2)\right)\cdot \left (\nabla\vec{u}+(\nabla
\vec{u})^t\right)\right).\nonumber
\end{eqnarray}
\par Linearizing eqs. (\ref{perturbaciones}) the following system
of equations is obtained
\begin{eqnarray}
&& \nabla \cdot \vec{u}=0, \nonumber\\
&& \partial _{t}\vec{u}=\alpha \theta g \vec{e}_3-\frac{1}{%
\rho _{o}}\nabla \left( \delta P\right) + L \vec{u},
\label{perturbaciones2} \\
&& \partial _{t}\theta =K\Delta \theta +\beta u_{z},\nonumber
\end{eqnarray}%
where displaying the operator $\displaystyle \tilde{L} \vec{u}=
\partial_z \left(\nu(z) \partial_z \vec{u} \right) +
\nu(z) \Delta' \vec{u} + \nu'(z) \nabla u_z$ and $\Delta'$ is the laplacian operator
in the variables $x$ and $y$.\par
 The system (\ref{perturbaciones2}) together with its
boundary conditions may be expressed in dimensionless form
defining new variables $\vec{r}'=\vec{r}/d $, $t'=K t/d^2$,
$\vec{u}'=d \vec{u}/K$, $p'= d^2 p/(\rho_o K \nu_o)$, $\theta' =
(\theta-T_1)/\beta d$. After rescaling the variables and dropping
the primes the system is left as follows,
\begin{eqnarray}
&& \nabla \cdot \vec{u}=0, \nonumber\\
&& \frac{1}{Pr}\partial _{t}\vec{u}=-\nabla \left( \delta P\right)
+ \tilde{L} \vec{u} + R \theta \vec{e}_3,
\label{perturbacionesadim} \\
&& \partial _{t}\theta =\Delta \theta + u_{z}.\nonumber
\end{eqnarray}
Here, the Rayleigh number is defined as $R=d^4 \alpha g \beta
/(\nu_0 K)$ and the Prandtl number is $ Pr = \nu_0 /K$, where
$\nu_0=\nu(-1/2)$.\par The dimensionless boundary conditions for
fields in (\ref{perturbacionesadim}) are
\begin{eqnarray}
&& \theta=u_x=u_y=u_z=0 \;{\rm on} \; z= -1/2,\;\theta=\partial_z
u_x =
\partial_z u_y = u_z =0 \;{\rm on} \; z= 1/2.\label{bcpert}
\end{eqnarray}
\subsection{Decomposition into normal modes}
As the domain is infinite in the plane, the eigenvalue problem can
be solved by using the normal modes method. For each mode a
separate variable solution can be found $\vec{S}(z)\cdot
f(x,y,t)$, where $f(x,y,t)$ can be expressed as a normal mode
$f(x,y,t)=e^{\sigma(\vec{k}) t+i( k_{x}x+k_{y}y )}$. Therefore we
look for solutions as follows,
\begin{equation}
\left(
\begin{array}{c}
u_x \\
u_y \\
u_z \\
\theta \\
\delta P \\
\end{array}
 \right) = \left(
\begin{tabular}{l}
$U_{x}(z)$ \\
$U_{y}(z)$ \\
$W(z)$ \\
$\Theta \left( z\right) $ \\
$\delta p(z)$%
\end{tabular}%
\right) \cdot e^{\sigma(\vec{k}) t+i\left( k_{x}x+k_{y}y\right) },
\label{solutions}
\end{equation}%
where $\sigma (\vec{k}) $ are the eigenvalues, that initially can
be complex numbers. It represents the growing rate of the normal
mode at the wave number $\left( k_{x},k_{y}\right) =\vec{k}$. If
the solution belongs to $L^2$ in $k_x,\,k_y$, the inverse Fourier
transform of the normal modes can be considered the solution of
the linear system.\par Introducing the solution (\ref{solutions})
into equations (\ref{perturbacionesadim}) we obtain the following
equations:\par \noindent
 $\bullet$ the continuity
equation,
\begin{equation}
i\left( k_{x}U_{x}+k_{y}U_{y}\right) +DW=0, \label{cont_modes}
\end{equation}%
$\bullet$ the Navier-Stokes equations,%
\begin{eqnarray}
\frac{\sigma}{Pr} U_x &=& - i k_x \delta p + \mathcal{P} U_x + i
k_x \nu'(z) W,\label{U_x_modes}\\
\frac{\sigma}{Pr} U_y &=& - i k_y \delta p + \mathcal{P} U_y + i
k_y \nu'(z) W,\label{U_y_modes}\\
\frac{\sigma}{Pr} W &=& - D (\delta p) + \mathcal{P} W + \nu'(z) D
W+R \Theta,\label{W_modes}
\end{eqnarray}
$\bullet$ the heat equation,%
\begin{equation}
\sigma \Theta =\left( D^{2}- k^{2}\right) \Theta +W,
\label{heat_modes}
\end{equation}%
where the operators $\mathcal{P} f=D(\nu Df) - \nu k^2 f$ and $D
f= df/dz$ and the wave number $k^2 = k_x^2 + k_y^2$ have been
introduced. \par The boundary conditions become
\begin{eqnarray}
\Theta= U_x=U_y=W=0 \;\mbox{on}\; z=-1/2;\;\Theta=DU_x=DU_y=W=0
\;\mbox{on}\; z=1/2. \label{bc_modes}
\end{eqnarray}

\begin{lemma}
If $(U_x,U_y,W,\Theta,\delta P)$, such that $(U_x,U_y,W)\in H^2$,
is a solution of problem (\ref{cont_modes}-\ref{heat_modes}) with
b.c. (\ref{bc_modes}) then it is a solution of the following
problem
\begin{eqnarray}
&& \frac{\sigma}{Pr}Z=\nu \left( z\right) \left[ \left(
D^{2}-\left\vert
k\right\vert ^{2}\right) Z\right] +\nu ^{\prime }\left( z\right) \cdot DZ, \label{eiga}\\
&& \sigma \Theta =\left( D^{2}-\left\vert k\right\vert
^{2}\right) \Theta + W, \label{eigb} \\
&& \frac{\sigma}{Pr}(D^2- |k|^2)W=-R |k|^2\Theta - |k|^2
\mathcal{P} W+ D(\mathcal{P}(DW))+|k|^2\nu''(z)W,\label{eigc}
\end{eqnarray}
\begin{equation}
Z=W=DW = \Theta =0,\;\;\mbox{on}\; z=- 1/2;\; Z=W=D^2W = \Theta
=0,\;\;\mbox{on}\; z= 1/2.\label{eigbc}
\end{equation}
Where $Z\left( z\right)=U_{y}\left( z\right) ik_{x}-U_{x}\left(
z\right) ik_{y}$, and the velocity functions $U_x$ and $U_y$ can
be calculated from $W$ and $Z$.
\end{lemma}
\begin{proof}
Pressure is perfectly determined by the velocity $(U_x,U_y,W)$ and
the temperature $\Theta$, so we proceed to eliminate the pressure
applying $(ik_x,ik_y,D)\times (U_x,U_y,W)$ into the Navier-Stokes
equations (\ref{U_x_modes}-\ref{W_modes}) hence
$ik_x\frac{\sigma}{Pr}U_y-ik_y\frac{\sigma}{Pr}U_x$ leads to
$(\ref{eiga})$.\par On the other hand, from the equations
(\ref{U_x_modes}-\ref{U_y_modes}) and together the continuity
equation (\ref{cont_modes}) we can obtain
\begin{equation}
-\frac{\sigma}{Pr}DW=|k|^2\delta P-\mathcal{P}(DW).
\end{equation} Hence one deduces the decoupled equation
\begin{equation}
\frac{\sigma}{Pr}|k|^2W=\mathcal{P}|k|^2W+R|k|^2\Theta-D\left [
\mathcal{P}(DW)-\frac{\sigma}{Pr}DW \right ]-|k|^2\nu''(z)W,
\end{equation}
which rewrites as (\ref{eigc}) and therefore the final system
(\ref{eiga}-\ref{eigc}) is obtained.
\par Finally, if we define the velocity functions
$U_{x}$, $U_{y}$ as,
\begin{equation}
U_{x}=ik_{x}\phi -ik_{y}\psi,\,\ U_{y}=ik_{y}\phi +ik_{x}\psi ,
\end{equation}%
using the continuity equation (\ref{cont_modes}) and knowing that
$Z\left( z\right)
=U_{y}\left( z\right) ik_{x}-U_{x}\left( z\right) ik_{y}$, we obtain that $%
\phi (z)$ and $\psi (z)$ are functions of $W\left( z\right) $ and
$Z(z)$ respectively,
\begin{equation}
\left\{
\begin{array}{c}
\displaystyle U_{x}\left( z\right) =i\left[ k_{x}\left( \frac{DW\left( z\right) }{k^{2}}%
\right) +k_{y}\left( \frac{Z(z)}{k^{2}}\right) \right] \\
\displaystyle U_{y}\left( z\right) =i\left[ k_{y}\left( \frac{DW\left( z\right) }{k^{2}}%
\right) -k_{x}\left( \frac{Z(z)}{k^{2}}\right) \right] %
\end{array}%
\right.
\end{equation}
so, we can reduce the system (\ref{cont_modes}-\ref{heat_modes})
in (\ref{eiga}-\ref{eigc}) without the terms with $U_x$ and $U_y$.
\end{proof}

\subsection{Real eigenvalues}
\bigskip
\begin{theorem} The eigenvalues for problem (\ref{eiga}-\ref{eigbc}) are
real.
\end{theorem}
\begin{proof}
Notice that, for any function $f\in
H_0^1([-\frac{1}{2},\frac{1}{2}])$, we have the identity
$$\int_{-\frac{1}{2}}^{\frac{1}{2}}\mathcal{P}f \overline{f}dz=
-k^2\int_{-\frac{1}{2}}^{\frac{1}{2}}\nu(z)\left [ {|d_hf|}^2
+{|f|}^2  \right ]dz$$ as well as the identity
$$\int_{-\frac{1}{2}}^{\frac{1}{2}}(d_h^2-1)f \overline{f}dz=
-\int_{-\frac{1}{2}}^{\frac{1}{2}}\left [ {|d_hf|}^2 +{|f|}^2
\right ]dz,$$ where $d_{h}=hD=D/k= d/dz/k$. We consider the
equations (\ref{eigb}-\ref{eigc}) and write them as follows,
\begin{eqnarray}
 \sigma \Theta (z) &=& k^{2}\left( d_{h}^{2}-1\right) \Theta
\left( z\right) + W\left( z\right),\label{heat1}\\
  -R\Theta (z)&=&\frac{\sigma}{Pr} \left( d_{h}^{2}-1\right)
W(z)-k^{2}d_{h}^{2}\left(\nu(z)d_h^2 W(z)\right)+ 2 k^2 d_h
(\nu(z) d_h W(z)) \nonumber\\
&& - k^2 \nu(z) W(z) - k^2 d_h (\nu(z) W(z)), \label{theta1}
\end{eqnarray}%
where we have cleared the temperature function and the derivatives $d^j_{h}=h_jD^j=D^j/k^j= d^j/dz^j/k^j$ .\\
\begin{equation}
\displaystyle \int W \overline{\Theta}dz=\sigma \int \left\vert
\Theta \right \vert ^{2}dz+k^{2} \int \left (\left\vert \Theta
\right\vert ^{2}+\left\vert d_{h}\Theta \right \vert ^{2}\right
)dz . \label{expres1}
\end{equation}
On the other hand, we consider the complex conjugate in the
equation (\ref{theta1}), we multiply it by $W(z)$ and integrate,
obtaining
\begin{eqnarray}
\displaystyle \int W\overline{\Theta }dz &=& \frac{\overline{\sigma }}{%
Pr R}\int \left( \left\vert W\right\vert ^{2}+ \left\vert d_h
W\right\vert ^{2}\right )dz +\nonumber \\ && \frac{k^2}{R}\int
\nu(z)\left(| W| ^{2}+
 2 |d_{h}W |^{2}+|d_h^2 W|^2 + d_h^2 (|W|^2)\right)dz.%
\label{expres2}
\end{eqnarray}%
We thus deduce
\begin{eqnarray}
\sigma\int |\Theta|^2dz+k^2\int(|\Theta|^2+ |d_h\Theta|^2)dz &=&
\frac{{\bar
\sigma}}{Pr R}\int(|W|^2+|d_hW|^2)dz+ \nonumber \\
&& \frac{k^2}{R}\int \nu(z)\left (|W|^2+2|d_h W|^2+|d_h^2W|^2+
d_h^2 (|W|^2)\right )dz.\nonumber\\ \label{equality}
\end{eqnarray}
Considering the imaginary part in relation (\ref{equality}) we
obtain
\begin{equation}\mbox{Im}\sigma \left(\int
|\Theta|^2dz+\frac{1}{PrR}\int(|W|^2+|d_hW|^2)dz
\right)=0.\end{equation} So, if $\sigma$ is not real, we deduce
that $\Theta$ and $W$ are zero, hence is a trivial solution. The
theorem is proved.\end{proof}

\subsection{Existence of the eigenvalues}
\bigskip
Before going into the proof of existence of eigenvalues, some
previous lemmas are necessary. In what follows, we shall make use
of the following inequalities:
\begin{lemma}
\label{Poincare}For $f\in H^1_0([-\frac{1}{2}, \frac{1}{2}])$, we
have the Poincar{\'e} inequality  as follows
\begin{eqnarray}
&&\int_{-\frac{1}{2}}^{\frac{1}{2}}|d_hf|^2dz\geq
\frac{\pi^2}{k^2}\int_{-\frac{1}{2}}^{\frac{1}{2}}|f|^2dz,
\label{inq1}\\
&&\int_{-\frac{1}{2}}^{\frac{1}{2}}\nu(z)|d_hf|^2dz\geq
\frac{\delta_1}{k^2}\int_{-\frac{1}{2}}^{\frac{1}{2}}\nu(z)|f|^2dz,
\label{inq2} \end{eqnarray} where $\delta_1$ is a constant
depending on $\nu(z)$.
\end{lemma}
\begin{proof}
The inequality (\ref{inq1}) is classical. It is obtained through
the minimization of $\int_{-\frac{1}{2}}^{\frac{1}{2}}|Df|^2dz$
under the constraint $\int_{-\frac{1}{2}}^{\frac{1}{2}}|f|^2dz$
for $f$ element of $H^1_0([-\frac12, \frac12])$, the solution of
this problem being given by the function $\cos\pi z$, hence the
inequality. On the other hand, we can deduce using (\ref{inq1})
and $\nu_{min}\leq \nu(z)\leq \nu_{max}$ the inequality as follows
\begin{equation}
\int_{\Omega}\nu(z)|d_hf(z)|^2dz\geq
\nu_{min}\int_{\Omega}|d_hf(z)|^2dz\geq
\frac{\pi^2\nu_{min}}{k^2}\int_{\Omega} |f(z)|^2dz\geq
\frac{\pi^2\nu_{min}}{k^2\nu_{max}}\int_{\Omega} \nu(z)|f(z)|^2dz.
\end{equation}
So Lemma \ref{Poincare} is proved.\end{proof}
\\ \\
Now we introduce ${\hat \sigma}= \sigma/k^2$.

\begin{lemma}
$\forall {\hat \sigma} \in \R$ there exists $k \in \R$ such that
$R({\hat \sigma}, k)
>0$.
\end{lemma}
\begin{proof} Considering $\Omega=\left [ -\frac{1}{2},\frac{1}{2} \right ]$ and
using the Lemma 3 and the H{\"o}lder inequalities we obtain from
the heat equation (\ref{eigb})
\begin{equation}
\left(\widehat{\sigma}+1+\frac{\pi^2}{k^2}\right)||\widetilde{\Theta}||_{L^2(\Omega)}
\leq ||W||_{L^2(\Omega)}, \label{desigualdad1}
\end{equation}
where $\widetilde{\Theta }=k^{2}\Theta$. From the decoupled
equation (\ref{eigc}),
\begin{equation}
\left (\frac{\widehat{\sigma}}{Pr} + \min_{[-1/2,1/2]}\nu(z)\left
(1+\frac{\delta_1}{k^2}\right ) + \min_{[-1/2,1/2]} d_h^2 (\nu(z))
\right )||W||_{H_0^1(\Omega)}^2\leq \widetilde{R}||\widetilde
{\Theta}||_{L^2(\Omega)}||W||_{L^2(\Omega)}, \label{desigualdad2}
\end{equation}
where $\widetilde{R}=R/k^{4}$. Hence we get the estimate
\begin{equation}
\left(\frac{\widehat{\sigma}}{Pr} +
\min_{[-1/2,1/2]}\nu(z)\left(1+\frac{\delta_1}{k^2}\right)+
\min_{[-1/2,1/2]} d_h^2 (\nu(z))\right)\left({\widehat
\sigma}+\left(1+\frac{\pi^2}{k^2}\right)\right)||W||_{H_0^1(\Omega)}^2\leq
{\widetilde R}||W||_{L^2(\Omega)}^2. \label{desigualdad3}
\end{equation}
We must thus deduce from (\ref{desigualdad3}) that if
$\widehat{\sigma}\geq 0$ then
\begin{equation}
\left( \min_{[-1/2,1/2]}\nu(z)
\left(1+\frac{\delta_1}{k^2}\right)+ \min_{[-1/2,1/2]} d_h^2
(\nu(z)) \right)
\left(1+\frac{\pi^2}{k^2}\right)||W||_{L^2(\Omega)}^2\leq
{\widetilde R}||W||_{L^2(\Omega)}^2, \label{desigualdad4}
\end{equation}
hence, using that $k>0$
$$R^*(k)=\left( \min_{[-1/2,1/2]}\nu(z)(k^2+\delta_1)+ k^2 \min_{[-1/2,1/2]} d_h^2 (\nu(z)) \right)
(k^2+\pi^2)\leq R(k),$$ so $R(k)>0$ because $d_h^2 (\nu(z)) >0$
$\forall z \in [-1/2,1/2]$. On the other hand, if $\sigma<0$ we
consider from (\ref{desigualdad3}) that
$$|\widehat{\sigma}|<min\left \{\min_{[-1/2,1/2]}\nu(z)\left (
1+\frac{\delta_1}{k^2}\right ), \min_{[-1/2,1/2]} d_h^2 (\nu(z)),
\mbox{ }\left ( 1+\frac{\pi^2}{k^2}  \right ) \right \}
$$ and we also obtain that $R(k)>0$.\end{proof}

Define the operator
$\mathcal{M}=(1-d_h^2+\widehat{\sigma})$. We have
\begin{lemma}\label{bcm}
Any solution of problem (\ref{eiga}-\ref{eigbc}) satisfies the boundary
conditions:$$\mathcal{M}^{1/2}
\widetilde{\Theta}(\pm\frac{1}{2})=0.$$
\end{lemma}
\begin{proof} First of all, we identify the eigenvalues and
eigenvectors of the operator
$\mathcal{M}=(1-d_h^2+\widehat{\sigma})$. For this purpose, we
solve
$$(1+\widehat{\sigma} - d_h^2)T=\lambda T$$
with boundary conditions $T(\pm\frac{1}{2})=0$. This rewrites
$$(d_h^2 + \lambda - (1+\widehat{\sigma}))\Theta=0.$$
As the equation $(d_h^2-\omega^2)u=0$ has not non-trivial solution
in $H^1_0$ for $\omega$ real, we have necessarily
$\lambda >(1+{\hat \sigma})$.\\
We introduce $\omega$ such that $\omega^2= \lambda -1 -
\widehat{\sigma}$. We obtain
$$T(x)=A\cos(\omega k x)+B\sin (\omega k x),$$
from which we deduce the condition to have a non zero non trivial
solution $\sin \omega k =0,$ hence $\omega k = n\pi.$ The
associated eigenfunction is $\cos(n\pi x)$ for even $n$ and is
$\sin(n\pi x)$ for odd $n$. We denote these functions by $T_n(x)$
and thanks to Fourier analysis, $L^2$ is described by a sum $\sum
a_n T_n$ where $\sum |a_n|^2<+\infty$ and $H^1_0$ is described by
$\sum a_n T_n$ with $\sum n^2|a_n|^2<+\infty$. We have
$\mathcal{M}(T_n)=
(\frac{n^2\pi^2}{k^2}+1+\widehat{\sigma})T_n=\lambda_n(\widehat{\sigma})T_n$.
Hence the operator $\mathcal{M}^{1/2}$ is a symmetric operator
with the eigenvalues $\mu_n(\widehat{\sigma})=
\sqrt{\lambda_n(\widehat{\sigma})}$ and for ${\tilde \Theta}\in
H^1_0([-\frac{1}{2}, \frac{1}{2}])$ we have $\mathcal{M}{\tilde
\Theta} =\sum a_n\sqrt{\mu_n({\hat \sigma})}T_n$.

As $\mathcal{M}({\hat \sigma}){\tilde \Theta}=W$ and $W\in H^2_0$,
$\mathcal{M}^{1/2}{\tilde \Theta}=\mathcal{M}^{-1/2}W$ and
$\mathcal{M}^{1/2}{\tilde \Theta} \in H^1_0$. Because $W=\sum w_n
T_n$ with $\sum n^4|w_n|^2<+\infty$ and $\sum
w_nT'_n(\pm\frac{d}{2})=0$. We deduce that
$\mathcal{M}^{-1/2}W=\sum \mu_n^{-1/2}w_nT_n$. This is an element
of $H^3$, hence we can compute the trace on the boundary, and as
the trace on the boundary of the eigenfunctions is zero, we have
$\mathcal{M}^{-1/2}W(\pm\frac{1}{2})=0$. Hence $\mathcal{M}^{1/2}
\widetilde{\Theta}(\pm\frac{1}{2})=0$
\end{proof}

\begin{theorem}
For all $k \in \R$ and for all $\widehat{\sigma}>0$, there exists
a function $R_{\perp}(k, \widehat{\sigma})$ satisfying
$R_{\perp}(k, \widehat{\sigma})\geq R^*(k)$ such that problem
(\ref{eiga}-\ref{eigc})with the bc (\ref{eigbc}) has a unique
solution $(Z,\Theta,W) \in (H_0^1)^2 \times H^2_0$, where $H^2_0$
are the functions of $H^2$ that fulfill the bc (\ref{eigbc}) for
$R=R_{\perp}(k, \widehat{\sigma})$. This is in particular true for
$\widehat{\sigma}=0$.
\end{theorem}
\begin{proof}
We consider the equations (\ref{eigb}-\ref{eigc}) and we write
them as follows:\begin{eqnarray}
&&\mathcal{M}(\widehat{\sigma})\widetilde{\Theta}(z)=W(z),
\label{eigb_homogeneizado}\\
&&\mathcal{L}(\widehat{\sigma})W(z)=\widetilde{R}(k)\cdot
\widetilde{\Theta}(z), \label{eigc_homogeneizado}
\end{eqnarray}
where the operators $\mathcal{M}$ and $\mathcal{L}$ are defined as
\begin{eqnarray}
\mathcal{M}&=&(1-d_h^2+\widehat{\sigma}), \\
\mathcal{L}&=& \left [ d_h\left (
Q-\frac{\widehat{\sigma}}{Pr}\right )d_h - \left (
Q-\frac{\widehat{\sigma}}{Pr}\right ) + d_h^2 (\nu(z))\right ]
\end{eqnarray}
 with
$Qf=\mathcal{P} f /k^2=d_{h}\left( \nu \left( z\right) \cdot
d_{h}f\right) -\nu \left( z\right) \cdot f$. \par The equations
(\ref{eigb_homogeneizado}-\ref{eigc_homogeneizado}) lead to
following eigenvalue problem
\begin{equation} \left\{
\begin{array}{c}
\displaystyle
L(\widehat{\sigma})\widetilde{\Theta}(z)=\widetilde{R}(k)\cdot
\widetilde{\Theta}(z)\\
\displaystyle \Theta(-1/2)=\Theta(1/2)=0 \label{eigen-problem}
\end{array}%
\right.
\end{equation} where the operator $L$ is not a self-adjoint
operator and is a composition expressed as follows
$$L=\mathcal{L}\mathcal{M}.$$ Our operator $L$ is not a
self-adjoint operator, but $\mathcal{L}$ and $\mathcal{M}$ are
 self-adjoint operators. In fact $\mathcal{M}$ is a positive operator,
\begin{eqnarray}
<\mathcal{M}\widetilde{\Theta}, \widetilde{\Theta}>=
(1+\widehat{\sigma})\int |\widetilde{\Theta}|^2 + \int |d_h
\widetilde{\Theta}|^2, \label{positive2}
\end{eqnarray}
so it is a symmetric operator. As $\mathcal{M}$ is self-adjoint
there is a orthogonal basis of $H_0^1$ with the eigenvalues
$\lambda_n \in \mathbf{R}$. Hence the operator $\mathcal{M}^{1/2}$
is a symmetric operator with eigenvalues $\mu_n=\sqrt{\lambda_n}
\in \mathbf{R}$. On the other hand $\mathcal{M}^{1/2}$ is a closed
operator because Sobolev embedding, $H^2\hookrightarrow H_0^1
\hookrightarrow L^2$. Therefore by Von Neumann theory
$\mathcal{M}^{1/2}$ is a diagonal self-adjoint operator. The
problem (\ref{eigen-problem}) is equivalent to the Sturm-Liouville
problem as follows
\begin{equation} \left\{
\begin{array}{c}
\displaystyle
H(\widehat{\sigma})\varphi(z)=\widetilde{R}(k)\cdot \varphi(z)\\
\displaystyle \varphi(-1/2)=\varphi(1/2)=0 \label{Sturm-Liouville}
\end{array}
\right.
\end{equation}
where the Dirichlet boundary conditions can be used thanks
to lemma \ref{bcm}, the unknown solution is
$\varphi=\mathcal{M}^{1/2}\widetilde{\Theta}$, and
$H=\mathcal{M}^{1/2}\mathcal{L}\mathcal{M}^{1/2}$ is a
self-adjoint operator because $\mathcal{M}^{1/2}$ and
$\mathcal{L}$ are self-adjoint operators. $\mathcal{L}$ is a
self-adjoint operator from $H^2_0$ to $H^{-2}$, hence $H$ is a
self-adjoint operator from $H^3_0$ to $H^{-3}$. Therefore
$\widetilde{R}(k)=0$ is not eigenvalue, so the Sturm-Liouville
problem
\begin{equation} \left\{
\begin{array}{c}
H(\widehat{\sigma})\varphi(z)=0\\
\varphi(-1/2)=\varphi(1/2)=0 \label{Sturm-Liouville0}
\end{array}
\right.
\end{equation}
has only the trivial solution as solution. Using the Freedholm
Alternative theorem we can deduce that there exists a function
$R_{\perp}(k)$, which is the smallest eigenvalue of the operator
$H$ with Dirichlet boundary conditions and the problem
(\ref{eigen-problem}) has a unique and not trivial solution
$\widetilde{\Theta}(z)$. Hence there exists a Green function
$G(z,\xi)$ such that
$$\Theta(z)=\frac{R(k)}{k^4}\int_{-\frac{1}{2}}^\frac{1}{2}G(z,\xi)\Theta(\xi)d\xi,$$
and the theorem is proved.
\end{proof}

\subsection{Exchange of stability}
\bigskip
\begin{theorem}
There exists a $3 \times 3$ matrix $M(R,k)$ such that the values
$R_o(k)$ for which $\sigma(R_o(k),k)=0$ fulfill the condition
det$(M(R_o(k),k))=0$.
\end{theorem}
\begin{proof}
For any $k \in \R$ we calculate the values of $R$ for which
$\sigma(R(k),k)=0$. From the equations (\ref{heat1}-\ref{theta1}),
we consider the heat and decoupled equations as follows,
\begin{eqnarray}
&& \left( D^{2}-k^{2}\right) \Theta + W  = \sigma \cdot \Theta,
 \label{eq1}\\
&& R k^{2}\Theta = D\left( \mathcal{P} -\frac{\sigma}{Pr} \right)
DW-k^{2}\left( \mathcal{P} -\frac{\sigma}{Pr} - \nu''(z) \right)
W, \label{eq2}
\end{eqnarray}%
where $\mathcal{P} f=D\left( \nu \left( z\right) \cdot Df\right)
-k^{2}\left( \nu \left( z\right) \cdot f\right)$. Homogenizing
equations (\ref{eq1}-\ref{eq2}) we obtain
\begin{eqnarray}
&& \left( d_{h}^{2}-1\right) \widetilde{\Theta }-\widehat{\sigma }\widetilde{%
\Theta }+W=0, \label{eq1hh}\\
&& \widetilde{R}\cdot \widetilde{\Theta }=d_{h}\left(
Q-\frac{\widehat{\sigma}}{Pr} \right) d_{h}W-\left(
Q-\frac{\widehat{\sigma}}{Pr} - d_h^2 (\nu(z)) \right) W,
\label{eq2hh}
\end{eqnarray}%
where $Qf=\mathcal{P} f /k^2$, $\widetilde{R}=R/k^{4}$ and
$\widetilde{\Theta }=k^{2}\Theta$.

Therefore, from the system (\ref{eq1hh}-\ref{eq2hh}), renaming the
variables as follows
\begin{eqnarray}
W_{o} &=&W, \\
W_{1} &=&d_{h}W=d_{h}W_{o},  \nonumber \\
W_{2} &=&\nu \left( z\right) \cdot d_{h}^{2}W=\nu \left( z\right)
\cdot d_{h}W_{1},  \nonumber \\
W_{3} &=&d_{h}W_{2}-2\nu \left( z\right) \cdot W_{1},  \nonumber \\
\Theta _{o} &=&\widetilde{\Theta } , \nonumber \\
\Theta _{1} &=&d_{h}\widetilde{\Theta }=d_{h}\Theta _{o},
\nonumber
\end{eqnarray}%
we obtain the system of ordinary differential equations for the
velocity and the temperature functions
\begin{eqnarray}
&& d_{h}\Theta _{1}=\left( 1+\widehat{\sigma } \right)
\cdot \Theta _{o}- W_{o}, \\
&& \widetilde{R}\cdot \Theta _{o}=d_{h}W_{3}+\left(
\frac{\widehat{\sigma }}{Pr}+\nu \left( z\right) + d_h^2 (\nu(z))
\right) \cdot W_{o}-\left( \frac{\widehat{\sigma }}{\nu \left(
z\right) Pr}\right) \cdot W_{2}.
\end{eqnarray}
These equations admit the following matrix formulation
\begin{equation}
d_{h}\left(
\begin{array}{c}
W_{o} \\
W_{1} \\
W_{2} \\
W_{3} \\
\Theta _{o} \\
\Theta _{1}%
\end{array}%
\right) =\left(
\begin{array}{cccccc}
0 & 1 & 0 & 0 & 0 & 0 \\
\displaystyle 0 & 0 & \displaystyle \frac{1}{\nu \left( z\right) } & 0 & 0 & 0 \\
0 & 2\nu \left( z\right) & 0 & 1 & 0 & 0 \\
\displaystyle -\left( \frac{\widehat{\sigma }}{Pr}+\nu \left( z\right) + d_h^2(\nu(z))\right) & 0 & \displaystyle \frac{\widehat{%
\sigma }}{\nu \left( z\right) Pr} & 0 & \widetilde{R} & 0 \\
0 & 0 & 0 & 0 & 0 & 1 \\
\displaystyle -1 & 0 & 0 & 0 & \displaystyle 1+\widehat{\sigma } & 0%
\end{array}%
\right) \cdot \left(
\begin{array}{c}
W_{o} \\
W_{1} \\
W_{2} \\
W_{3} \\
\Theta _{o} \\
\Theta _{1}%
\end{array}%
\right)\label{system}
\end{equation}%
with boundary conditions
\begin{equation}
\left\{
\begin{array}{cc}
\displaystyle W_{o}\left( z= - \frac{1}{2}\right) =0 & \displaystyle W_{o}\left( z=  \frac{1}{2}\right) =0 \\
\displaystyle W_{1}\left( z= - \frac{1}{2}\right) =0 & \displaystyle W_{2}\left( z= \frac{1}{2}\right) =0\\
\displaystyle \Theta _{o}\left( z=- \frac{1}{2}\right) =0 &
\displaystyle \Theta _{o}\left( z= \frac{1}{2}\right) =0
\end{array}%
\right\}  \label{bc}
\end{equation}
In order to solve this problem we first consider the initial
conditions on $z=-1/2$, where initial conditions for
$W_2,\, W_3$ and $\Theta_1$ are deduced from the equations,
\begin{eqnarray}
W_{2}\left( z=-\frac{1}{2}\right) &=&\nu \left(
-\frac{1}{2}\right) \cdot
d_{h}W_{1}\left( -\frac{1}{2}\right), \\
W_{3}\left( z=-\frac{1}{2}\right) &=& d_{h}W_{2}\left(
-\frac{1}{2}\right),\\
\Theta _{1}\left( z=-\frac{1}{2}\right) &=& d_{h}\Theta _{o}\left(
-\frac{1}{2}\right).
\end{eqnarray}
As we do not know the values of $d_h W_1(-1/2)$, $d_h W_2(-1/2)$
and $d_h \Theta_o(-1/2)$ such that the boundary conditions on
$1/2$ are fulfilled, we consider the following general initial
conditions
\begin{equation}
\left\{
\begin{array}{c}
\displaystyle W_{o}\left( z=-\frac{1}{2}\right) =0 \\
\displaystyle W_{1}\left( z=-\frac{1}{2}\right) =0\\
\displaystyle W_{2}\left( z=-\frac{1}{2}\right) =\nu \left( -\frac{1}{2}\right) \frac{1}{k}%
A \\
\displaystyle W_{3}\left( z=-\frac{1}{2}\right) =\frac{1}{k}B \\
\displaystyle \Theta _{o}\left( z=-\frac{1}{2}\right) =0 \\
\displaystyle \Theta _{1}\left( z=-\frac{1}{2}\right) =\frac{1}{k}C%
\end{array}%
\right\} \label{bc1}
\end{equation}%
and we solve (\ref{system}) with bc (\ref{bc1}) for three
different sets of values of $(A,B,C)$: (1,0,0), (0,1,0) and
(0,0,1). We call the respective solutions $\vec{W1}$, $\vec{W2}$
and $\vec{W3}$. The general solution can be written as a linear
combination of the form
\begin{equation}
\vec{W}\left( z\right) =\lambda \cdot \vec{W1}%
\left( z\right) +\mu \cdot \vec{W2}\left( z\right) +\omega \cdot
\vec{W3}\left( z\right) ,
\end{equation}%
where%
\begin{equation}
\vec{Wi}\left( z\right) =\left( Wi_{o},Wi_{1},Wi_{2},Wi_{3},\Theta
i_{o},\Theta i_{1}\right) \left( z\right) ,
\end{equation}%
We look for a solution of (\ref{system}) with bc (\ref{bc}),
therefore the solution $\vec{W}$ must verify the boundary
conditions on $z=1/2$
\begin{equation}
\left\{
\begin{array}{c}
\displaystyle W_{o}\left( \frac{1}{2}\right) =\lambda \cdot W1_{o}\left( \frac{1}{2}%
\right) +\mu \cdot W2_{o}\left( \frac{1}{2}\right) +\omega \cdot
W3_{o}\left( \frac{1}{2}\right) =0 \\
\displaystyle W_{2}\left( \frac{1}{2}\right) =\lambda \cdot W1_{2}\left( \frac{1}{2}%
\right) +\mu \cdot W2_{2}\left( \frac{1}{2}\right) +\omega \cdot
W3_{2}\left( \frac{1}{2}\right) =0 \\
\displaystyle \Theta _{o}\left( \frac{1}{2}\right) =\lambda \cdot \Theta1_{o}\left( \frac{%
1}{2}\right) +\mu \cdot \Theta2_{o}\left( \frac{1}{2}\right)
+\omega \cdot \Theta3_{o}\left( \frac{1}{2}\right) =0
\end{array}
\right. \label{system1}
\end{equation}%
This system can be expressed as
\begin{equation}
M(R,k) \cdot \left(
\begin{array}{c}
\lambda \\
\mu \\
\omega%
\end{array}%
\right) =\left(
\begin{array}{c}
0 \\
0 \\
0%
\end{array}%
\right) , \label{lineal}
\end{equation}%
where the matrix $M(R,k)$ is defined by
\begin{equation} M(R,k) =\left(
\begin{array}{ccc}
\displaystyle W1_{o}\left( \frac{1}{2}\right) & \displaystyle
W2_{o}\left( \frac{1}{2}\right) &
\displaystyle W3_{o}\left( \frac{1}{2}\right) \\
\displaystyle W1_{2}\left( \frac{1}{2}\right) & \displaystyle
W2_{2}\left( \frac{1}{2}\right) &
\displaystyle W3_{2}\left( \frac{1}{2}\right) \\
\displaystyle \Theta 1_{o}\left( \frac{1}{2}\right) &
\displaystyle \Theta 2_{o}\left( \frac{1}{2}\right)
& \displaystyle \Theta 3_{o}\left( \frac{1}{2}\right)%
\end{array}
\right).
\end{equation}
Taking $\widehat{\sigma}=0$, problem (\ref{system}-\ref{bc}) has
solution, therefore the linear system (\ref{lineal}) has a
solution, i.e., there is a value of $\widetilde{R}$,
$\widetilde{R}_o(k)$, such that the system (\ref{system1}) has a
solution, i.e., there exist a value $\widetilde{R}_o(k)$ such that
det$(M(\widetilde{R}_o(k),k))=0$. Figure 2 shows this function
$F(R)$=det$(M(\widetilde{R},k))$, in the case of exponential
dependence of viscosity on temperature, for $\gamma= 3 \cdot 10^4$
and a fixed value of $k$, $k=2.54$, crossing the axis at $\widetilde{R}_o(k)$.%
\end{proof}
\begin{corollary}
For any $k \in \R$ there exists values of $\widetilde{R}$ for
which $\widehat{\sigma} >0$ and values of $\widetilde{R}$ for
which $\widehat{\sigma} <0$. Therefore, the exchange of stability
holds.
\end{corollary}
\begin{proof}
If we consider $\widehat{\sigma} >0$, problem
(\ref{system}-\ref{bc}) has solution, therefore the linear system
(\ref{lineal}) has a solution, i.e., there is a value of
$\widetilde{R}$, $\widetilde{R}^+(k)$, such that the system
(\ref{system1}) has a solution, i.e., there exist a value
$\widetilde{R}^+(k)$ such that det$(M(\widetilde{R}^+(k),k))=0$.
And the same applies to $\widehat{\sigma}<0 $.\end{proof}\par From
here the marginal velocity and temperature fields are easily
obtained. For $\widetilde{R}_{o}(k)$, the rank of the matrix is
less than three and the system to solve is
\begin{equation}
\left\{
\begin{array}{c}
\displaystyle \lambda \cdot W1_{o}\left( \frac{1}{2}\right) +\mu \cdot W2_{o}\left( \frac{1%
}{2}\right) =-\omega \cdot W3_{o}\left( \frac{1}{2}\right) \\
\displaystyle \lambda \cdot \Theta 1 _{o}\left( \frac{1}{2}\right)
+\mu \cdot \Theta 2
_{o} \left( \frac{1}{2}\right) =-\omega \cdot \Theta3_{o} \left( \frac{1}{2}%
\right)%
\end{array}%
\right.
\end{equation}%
with the free parameter $\omega $. The solution of this system is
the growing perturbation after the bifurcation.\par
For all $k \in
\R$ we calculate $\widetilde{R}_o(k)$, this function has a
minimum, i.e., there is $k_c$ such that $\widetilde{R}_o(k_c)=\min
\left\{\widetilde{R}_{o}(k),\,k \in \R \right\}$. The real $k_c$ is the
critical wave number and $\widetilde{R}_c=\widetilde{R}_o(k_c)$ is
the critical Rayleigh number. As $\widetilde{R}$ increases this is
the first value of $k$ for which an eigenvalue becomes positive.
The corresponding mode grows and the conductive state becomes
unstable. In fact, for any $\widetilde{R} <\widetilde{R}_{c}$, the
conductive solution is stable, it looses stability at
$\widetilde{R}_c$ and it is unstable for $\widetilde{R} >
\widetilde{R}_c$. The curve $\widetilde{R}_{o}(k)$ is the marginal
stability curve which is calculated in the next section for some
values of the parameters.

\section{Numerical Results}

In this section the viscosity is assumed to have an exponential
dependence on temperature $T$,
\begin{equation}
\nu (T)= \nu_{o} \cdot \exp (-\gamma (T-T_1)),
\end{equation}
where $\nu_0$ is the viscosity at temperature $T_1$ and $\gamma$
the exponential rate. Figure 3 shows two different viscosity
profiles for $\gamma =10$ and $\gamma=3 \cdot 10^{4}$, that
correspond to almost constant viscosity and strong variable
viscosity, respectively.


First we have calculated the critical values $R_o(k)$ for each $k$
looking for the first root of det$(M(R,k))$.  The marginal
stability curves for $\gamma=10$ and $\gamma=3 \cdot 10^4$ can be
seen in figure 4. The minima in the critical wave numbers $k_{c}=
2.68$ and $R_c=1082.90$ for $\gamma=10$ and $k_{c}= 2.16$ and
$R_c=73.51$ for $\gamma= 3 \cdot 10^4$ indicates the critical
thresholds for the first bifurcation. In table I the critical
Rayleigh and wave numbers for several values of $\gamma$ are
displayed. The corresponding critical Rayleigh number, $R_{c}$,
decreases when $\gamma$ increases. Therefore a higher variation of
the exponential rate in viscosity favors the instability. The
critical wave number also depends on the viscosity factor
$\gamma$. It decreases when $\gamma$ increases until $\gamma=5
\cdot 10^4$ where it increases again. As $\gamma$ tends to zero,
$\nu(z) \rightarrow \nu_0$ the results tend to the ones with
constant viscosity, $k_{c}=2.68$ and $R_{c}=1100.65$. Figure 5
shows the corresponding velocity functions of the perturbation
field for two cases: almost constant ($\gamma=10$) and strongly
variable viscosity ($\gamma= 3 \cdot 10^4$), for different
Rayleigh numbers, below the critical one, at the critical
threshold and above it. In the variable viscosity case the
velocity is larger in the region where viscosity is smaller, while
in the constant viscosity case velocity is more distributed along
the cell.
Results are similar to those in ref.
\cite{severin} where a linear approximation to the exponential
dependence is considered.

\section{Conclusions}

We have studied the thermoconvective instability problem in an
infinite layer under the perspective of viscosity as an
exponential function of the temperature. We have obtained the
conductive solution and we have studied its linear stability
analysis. We  have demonstrated that the eigenvalues are real and
the existence of eigenvalues and bifurcation.

With appropriate changes of variable we have considerably
simplified the expressions to obtain a system of ordinary
differential equations which is numerically manageable. We have
obtained a practical condition which permits easily to calculate
the critical bifurcation parameters.

Finally, we have obtained the stability curves for different
values of the exponential rate, $\gamma$. As the exponential rates
decrease the critical Rayleigh number increases. So a higher
variation of the exponential rate in viscosity favors the
instability. There is a non monotone variation of the critical
wave number with the exponential rate. The vertical velocity for
the growing mode exhibits more movement in the region where the
viscosity is lower.

\section*{Acknowledgments}

This work was partially supported by the Research Grant MEC
(Spanish Government) MTM2006-14843-C02-01 and CCYT (Junta de
Comunidades de Castilla-La Mancha) PAI08-0269-1261, which include
RDEF funds.

\newpage
\bigskip
\noindent Table captions

\bigskip

\noindent Table I

\noindent Critical wave and Rayleigh numbers ($k_c$,$R_c$) for
different values of $\gamma$.

\bigskip

\noindent Figure captions

\noindent Figure 1

\noindent Problem set-up.

\bigskip

\noindent Figure 2

\noindent $F(R)$=det$(M(R,k=2.54),k=2.54)$ for $\gamma= 3 \cdot
10^4$; this plot represents a transcritical bifurcation.

\bigskip

\noindent Figure 3

\noindent Viscosity profiles $\nu(z)$ for $\gamma=3 \cdot 10^4$
(dashed line) and $\gamma=10$ (solid line).

\bigskip

\noindent Figure 4

\noindent Marginal stability curves for $\gamma=3 \cdot 10^4$
(dashed line) and $\gamma=10$ (solid line).

\bigskip

\noindent Figure 5

\noindent a) Velocity functions $W$ of the growing mode for
different values of the Rayleigh number for $\gamma =10$ and
$k=2.68$; b) velocity functions $W$ of the growing mode for
different values of the Rayleigh number for $\gamma =3 \cdot 10^4$
and $k=2.16$.

\bigskip

\bigskip \newpage

\bigskip
\noindent
\begin{center}
Table I \\
$$\begin{tabular}[b]{|l|l|} \hline
$\gamma$&($k_c$,$R_{c}$)\\
\hline
$10^{-4}$&(2.68,1100.65) \\
10&(2.68,1082.90) \\
$10^3$&(2.18,153.66) \\
$3 \cdot 10^4$&(2.16,73.51) \\
$5 \cdot 10^4$&(2.42,50.62) \\
 \hline
\end{tabular}$$
\end{center}

\bigskip
\noindent
\begin{center}
\textit{Figure 1.}\\
\begin{tabular}{c}
\epsfig{file=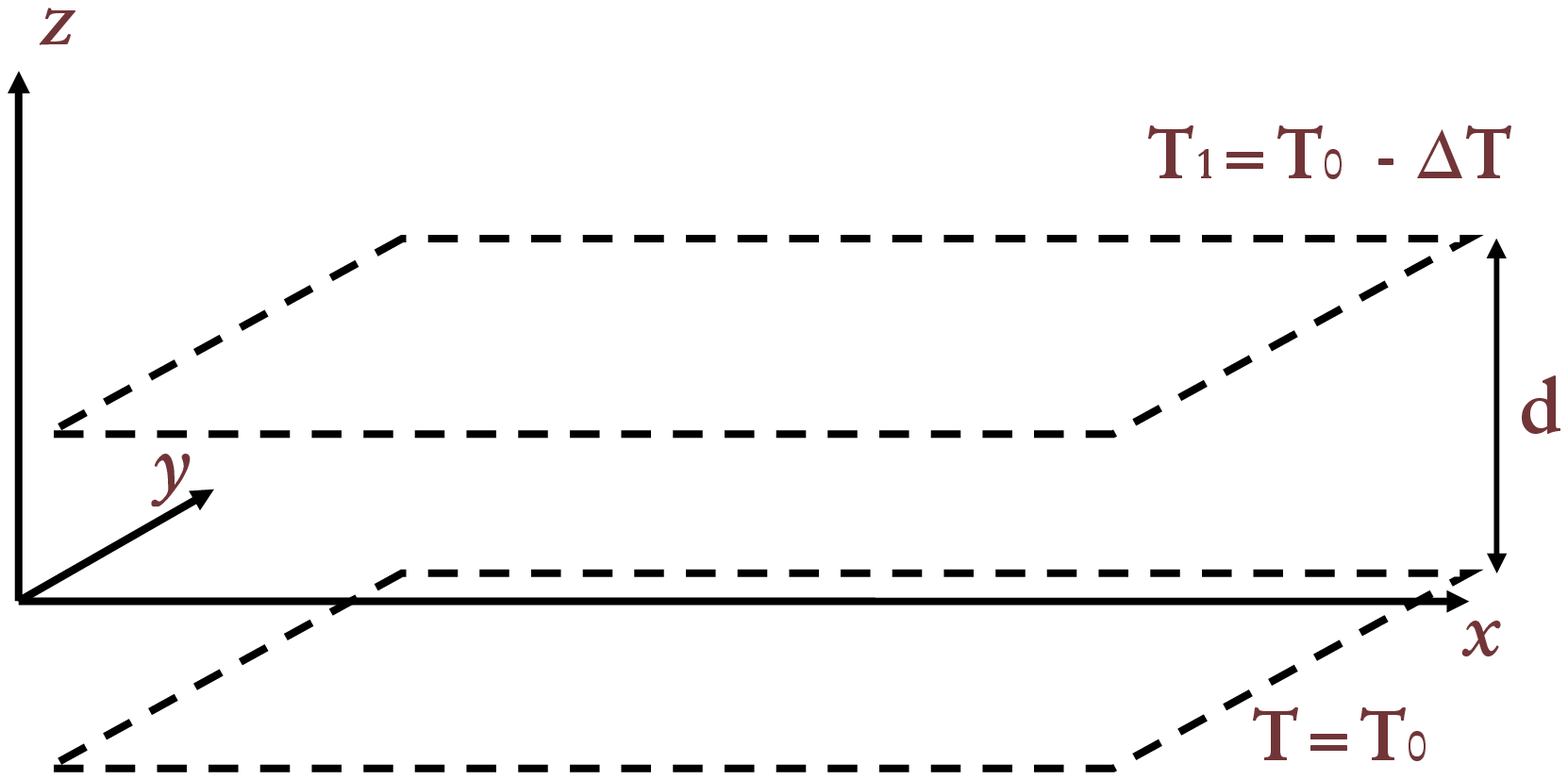,width=0.60\textwidth} \\
\end{tabular}
\end{center}

\bigskip
\noindent
\begin{center}
\textit{Figure 2.}\\
\begin{tabular}{c}
\epsfig{file=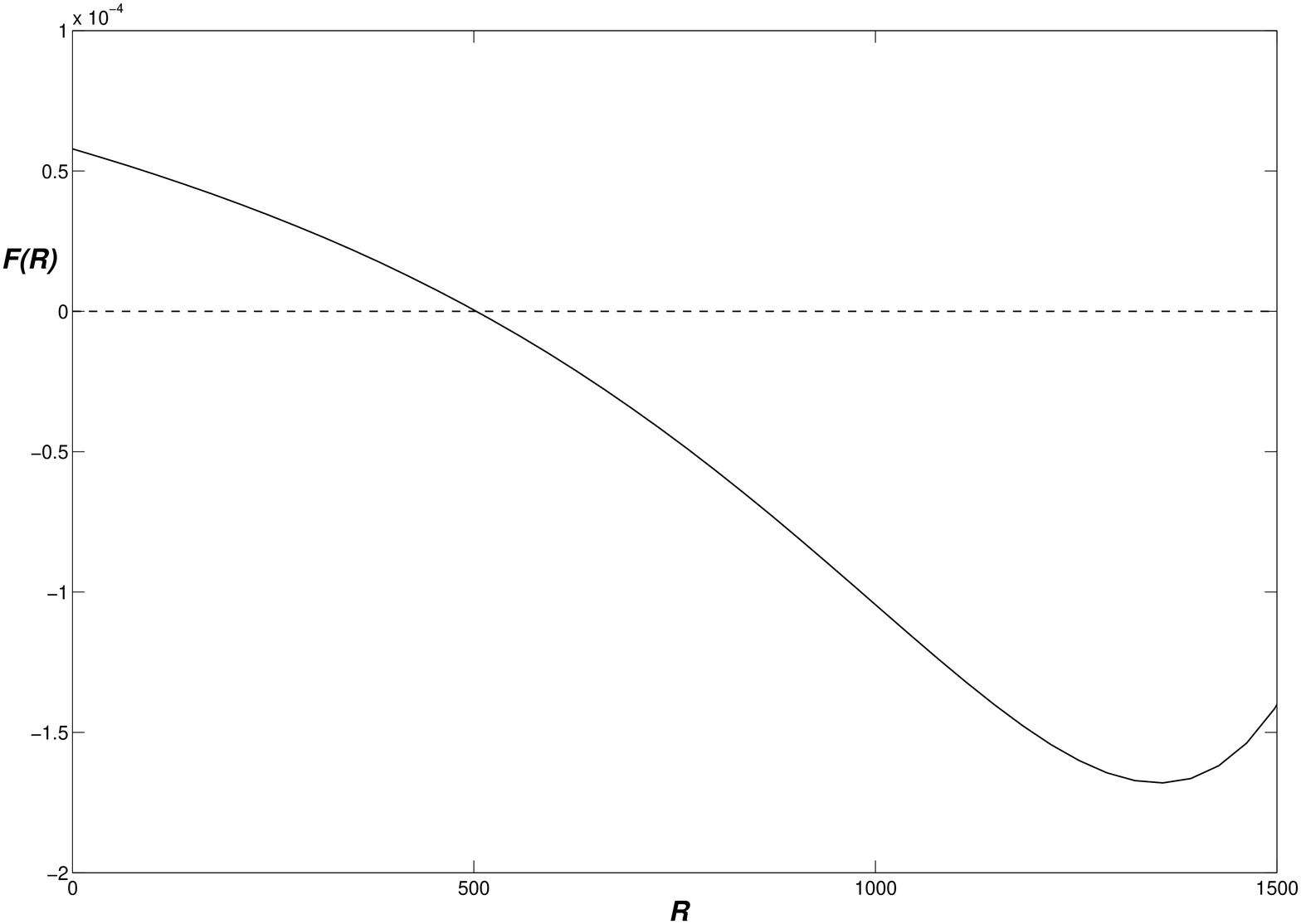,width=0.60\textwidth} \\
\end{tabular}
\end{center}

\bigskip
\noindent
\begin{center}
\textit{Figure 3.}\\
\begin{tabular}{c}
\epsfig{file=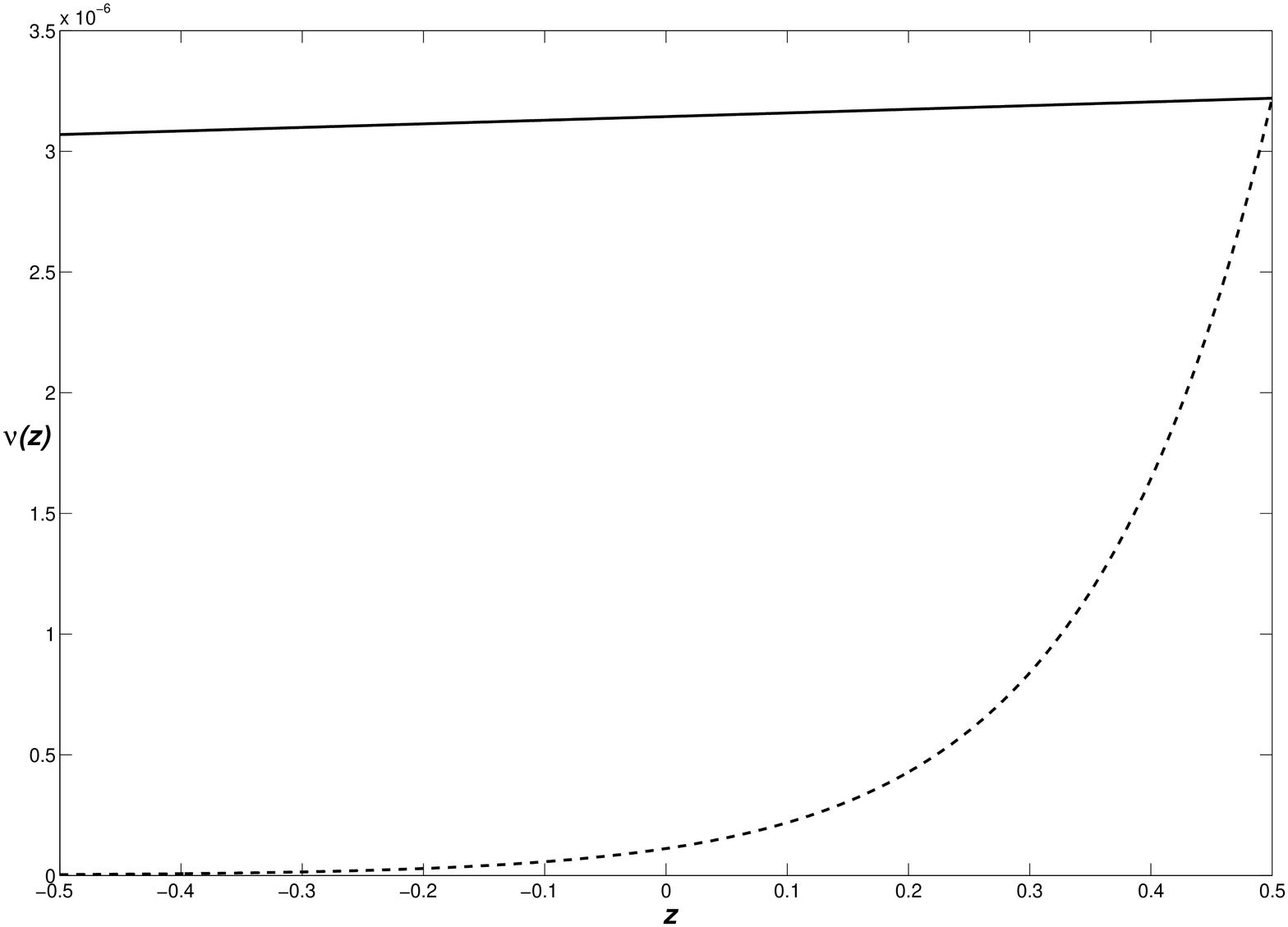,width=0.60\textwidth} \\
\end{tabular}
\end{center}

\bigskip
\noindent
\begin{center}
\textit{Figure 4.}\\
\begin{tabular}{c}
\epsfig{file=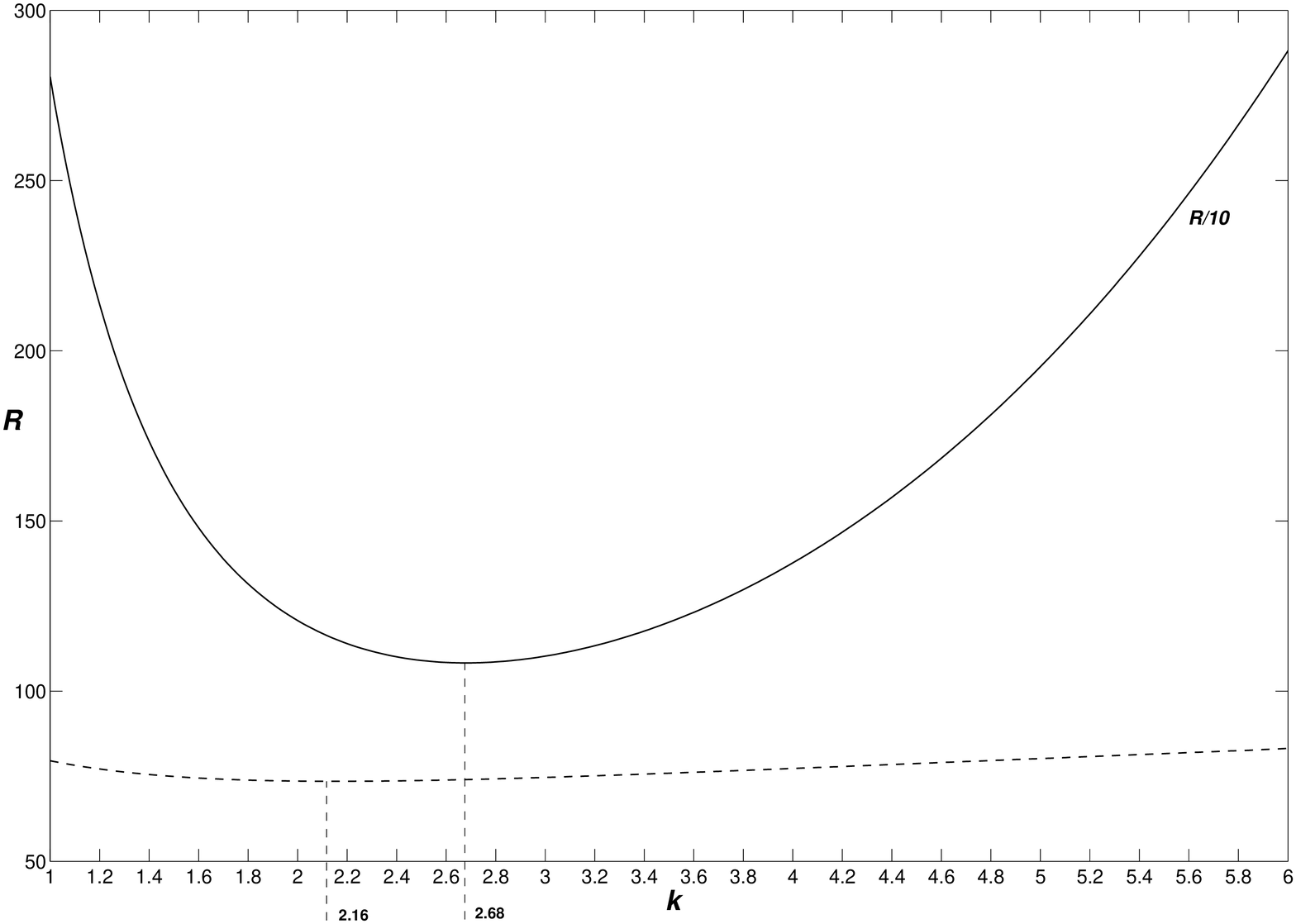,width=0.60\textwidth} \\
\end{tabular}
\end{center}

\bigskip
\noindent
\begin{center}
\textit{Figure 5.}\\
\begin{tabular}{c}
\epsfig{file=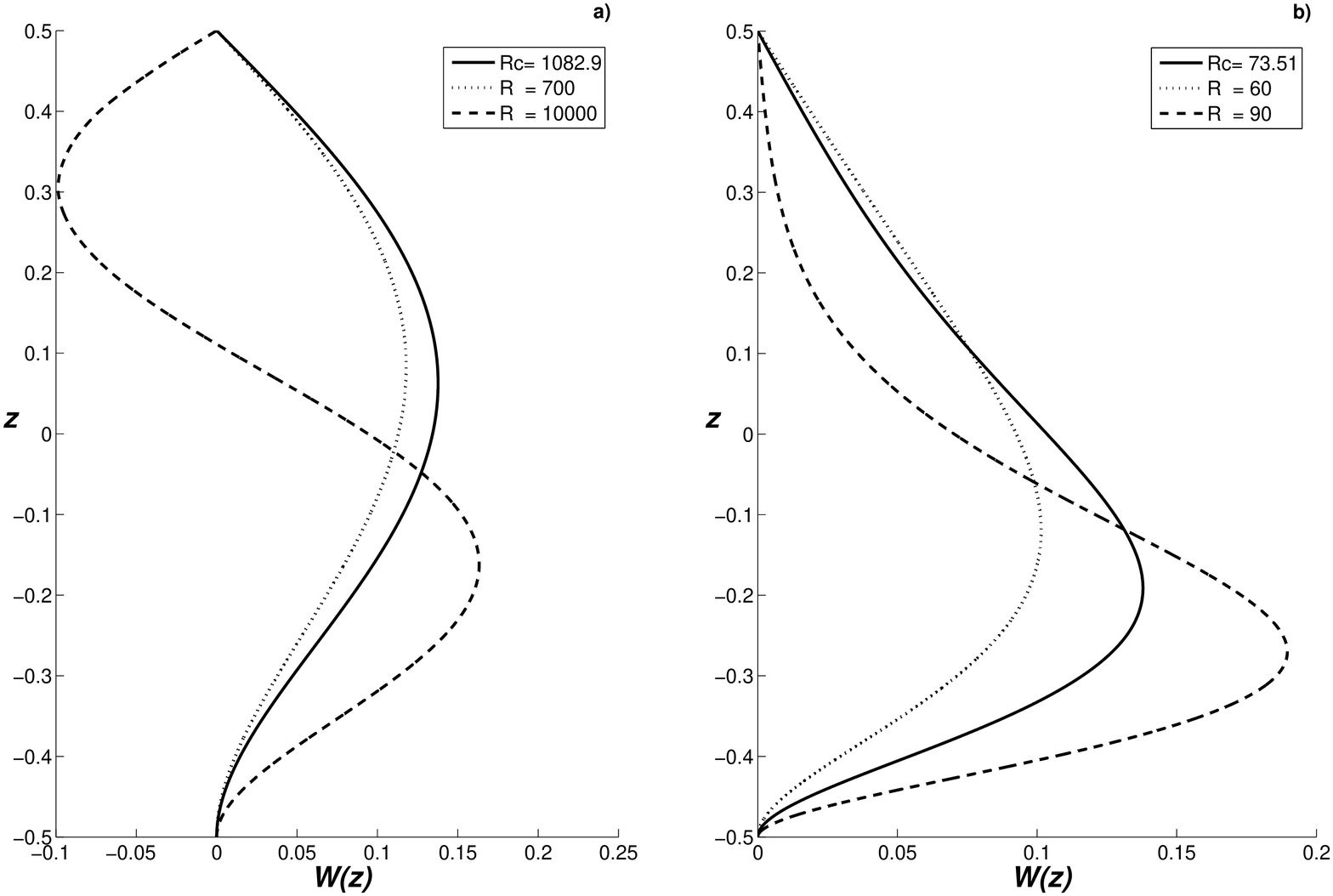,width=0.60\textwidth} \\
\end{tabular}
\end{center}

\end{document}